# A Variational Framework for Nonlinear Chemical Thermodynamics Employing the Maximum Energy Dissipation Principle

**Adam Moroz**

*Engineering Department, Faculty of Technology, De Montfort University, Leicester, UK*

The maximum energy dissipation principle is employed to nonlinear chemical thermodynamics in terms of distance variable (generalized displacement) from the global equilibrium, applying the optimal control interpretation to develop a variational formulation. The cost-like functional was chosen to support the suggestion that such a formulation corresponds to the maximum energy dissipation principle. Using this approach, the variational framework was proposed for a nonlinear chemical thermodynamics, including general cooperative kinetics model. The formulation is in good agreement with standard linear non-equilibrium chemical thermodynamics.

**Keywords:** variational approach, non-equilibrium thermodynamics, optimal control, cooperativity

## Introduction

Cooperative effects are of great interest throughout chemistry and physics, biology, information technology and cognition. Recent studies in a wide range of diverse fields c.f. submolecular cooperativity in multi-state protein unfolding and refolding,[1] cooperativity and specificity in enzyme kinetics,[2] effects of bond cooperativity in thermodynamics of hydrogen-bonded fluids,[3] cooperativity in protein motion,[4] kinetics of cooperativity-based drug delivery systems,[5] cooperativity between cell contractility and adhesion,[6] collective dynamics of bacteria,[7] cooperativity effects of clusters in some spirits,[8] cooperativity and hydrogen bonding network in water clusters,[9] structural cooperativity during protein unfolding,[10] cooperative effects in stabilization of the H-bonding in isoguanine trimers,[11] cooperative hydrogen bonding in amides and peptides,[12] cooperative effects that constitute the basic mechanism for the spin crossover transition,[13] cooperative behaviour in dynamic ferromagnetic cores,[14] synergetical effects,[15] superconductivity,[16] quantum phase transitions in dissipative systems,[16] quantum cooperative effects leading to superradiance,[18,19] cooperative enhancement of two-photon absorption in dendrimers,[20] social networks[21] and cognition[22] can give some idea of the extent of these phenomena. Cooperative effects are coupled phenomena, which are more common and can take place in linear and nonlinear region of thermodynamic branch. [23]

All these cooperative effects in chemistry and biology and collective effects in physics have a similar kinetical manifestation from the phenomenological thermodynamic perspective as the nonlinear and dissipative effects in the systems removed far from global equilibrium[15]. However, this similarity is not incorporated into the common high-level formalism, which exists separately only in physics. In physics this is the Lagrange/Hamilton formalism based on the variational approach and we can expect that its extension into other areas can help to narrow the gap between physics/chemistry and other fields mentioned above.

From an energy transformation (thermodynamic) perspective, the variational approach is related to some extreme principles. When considering phenomenological thermodynamic aspects of nonlinear phenomena we can utilize the maximum energy dissipation (MED) principle,[24-32] related to maximum entropy production (MEP) principle,[33,34] which in recent decades has been shown to rival the Prigogine minimum entropy production/energy dissipation principle[35-36] that is limited to steady state conditions.[30] Recently further progress was made in understanding that MEP/MED principles are more general[30,31] and can be directly related to the least action principle.[32] One can expect that the successful application of the variational approach could be a good interexchange for conceptual understanding and descriptions of cooperative phenomena in physics and chemistry on one side with biology and biokinetics, on the other. Consequently, the variational approach is closely related to optimal control methods[37] (e.g. the Pontryagin maximum principle[38]) – which have extensive applications in modern biokinetics, biochemistry and biotechnology. An extension of the understanding of the extreme properties in these processes may well be interesting for many modern bioengineering fields, including biochemical network design and optimization.

In former study[32] we employed an optimal control (OC) interpretation to the variational approach to formulate the MED principle for nonlinear chemical thermodynamics. In this letter we propose an outline for the further approximations of nonlinear (cooperative) kinetics, following our approach[32] together with the generalization of the linear model.

## Preliminaries

Formerly we described a based on dynamic optimal control variational formulation for the maximum energy dissipation principle in non-linear chemical kinetics, employing the positive definite thermodynamic potential and the positive definite dissipation function constructing thermodynamic Lagrangian. The direct optimal control interpretation of this choice was also engaged by applying the optimal control formulation of classical mechanics[32] as an example, where the control variables appeared as artificial variables in the equation

$$\dot{q}_i = u_i\,,\; i = 1,...N\,, \qquad (1)$$

where, the generalized velocities $\dot{q}_i$ become formally new variables $u_i$, i.e. control variables. In these terms the corresponding classic mechanical action becomes

$$S=\int_{t_1}^{t_2}\Lambda(u_i,q_i)dt=\int_{t_1}^{t_2}(T(u_i)-U(q_i))dt\rightarrow extr\,,$$

$$q_i(t_1)=q_{i1},\ q_i(t_2)=q_{i2}\ . \quad (2)$$

Here $S$ is the action, $\Lambda$ is the Lagrangian, $q_i$ are generalized coordinates, $U$ is the potential term, $T$ is the kinetic term, $t$ is time, $t_1$ and $t_2$ is the time interval, the control variables $u_i$ should be considered as having no restrictions. In this case we obtain a classical dynamic Lagrange optimal control problem.[32] It is well known that a powerful technique to solve the problem of eq 1-2 is based on the Pontryagin maximum principle.[37,38]

However, as illustrated recently[32] the case of the OC formulation of the classical mechanics problem, the interpretation of the Lagrangian $\Lambda$ from eq 2 seems to be cost-like and it is quite difficult to interpret the negatively defined term $U(q)$ from the OC cost-explicit perspective. Conversely, one can see that if this term $U(q)$ is positively defined it could simply be interpreted from OC in a cost-like manner as the energetical penalty for being in an non-equilibrium state.[32]

It was also stressed that in the OC formulation of mechanics, the control variables $u_i$ have less real meaning, the formulation seems ambiguous. However, in the case of a nonlinear system, when we are seeking a variational formulation, and a dynamical system is already formulated, this method might be seen as more useful and can be considered as a way to build a pure variational problem.

## Results and discussion

**General case of an OC-like problem.** In contrast to the mechanical example (eq 1-2), in a nonlinear system case extending eq 1, particularly characteristic for biochemical and biological kinetics, the systems certainly are nonlinear and we can be write

$$\dot{\xi}_i=f_i(\xi_1,...\xi_N,k_1...,k_j,...k_M)\,,\ \xi_i(t_0)=\xi_{i0} \quad (3)$$

here and later on, $\xi_i$ are generalized displacements from equilibrium, which are the phenomenological state variables characterizing the extent from the equilibrium (extent coordinates in chemical thermodynamics) and $k_j$ are the rate constants. Formulating a mechanical-like variational approach using eq 3 is far more difficult than for the mechanical processes described by eq 1, and our approach in employing optimal control, which seemed oversimplified at first for the mechanical case, can be much more useful in non-linear cases like in chemical kinetics or biological/biochemical kinetics. One possible method could be as follows.

Let us suggest that it is possible to vary the rate constants $k_j$ (as in a biochemical network) then we can treat these constants as the control variables. In this case we can also formally define an energetic cost (energetic loss for the deregulation, for deviation from an optimal value in the case of the metabolic networks)[32] for the rate constant deviations. This means that the optimal control of the overall kinetic process can be considered as controlled by the rate constants (not just by the rate values as it was in mechanical case eq 1). Then the system (eq 3) can be rewritten as

$$\dot{\xi}_i=f_i(\xi_1,...\xi_N,u_1...u_N)\,,\ \xi_i(t_0)=\xi_{i0}\,,\ i=1,...N \quad (4)$$

where $\xi_i$ are generalized displacements from equilibrium (extent coordinates in chemical thermodynamics), and $u_l$ are formal controls (considering a simplest case, $M=N$ in eq. 3). Following our approach[32] we can consider this system as the dynamical constraints, which have a more complicated relationship, than is the case in classical mechanics (eq 1). Then we can generalize the problem (eq 1-2) and make it more specific to nonlinear chemical thermodynamics, taking into account the fact that the chemical relaxation processes represent an open end problem. Let us choose the minimization functional as

$$S=\int_{t_0}^{\tau}\Lambda(u_i,\xi_i)dt=\int_{t_0}^{\tau}(\Phi(u_i)+\Psi(\xi_i))dt\rightarrow\min \quad (5)$$

following approach,[32] with an open-end, subject to a dynamical system eq 4, here and later with a fixed initial time $t_0$, unspecified final time $\tau$, and a fixed target state $\xi_i=0$, $\Phi$ is the dissipative function (the quadratic function in the vicinity of global equilibrium[24]), $\Psi$ is the thermodynamic potential. We shall consider this optimal control problem having no formal restrictions on the control variables $u_i$ or on the state variables $\xi_i$. Then applying the Pontryagin maximum principle one can construct the OC Hamiltonian

$$H(\xi_i,u_i,p_i)=-\Phi(u_i)-\Psi(\xi_i)+\sum_{i=1}^{N}p_if_i\,,\ i=1,...N\,, \quad (6)$$

and using the Pontryagin conditions[38] we can find optimal solutions. Since the final time $\tau$ is free and all $\xi_i$ at the unspecified time are equal to zero, no terminal condition is specified. Because of this at each point of optimal trajectory (*) the Hamiltonian (eq 6) is equal to zero: $H(\xi_i^*,u_i^*,p_i^*)=0$, (the additional demand of the Pontryagin maximum principle for an open-end optimal control problem at so-called natural boundary conditions). In such a direct OC formulation, the control variables (rate constants of a process) have a much more real meaning, in comparison to the artificial appearance of $u_i$ when the classic mechanical problem was written in OC terms, (eq 1-2). The problem in eq 4-5 can be reformulated as a variational problem when the control variables are singled out.

**Lagrange formulation.** Now, to formulate a pure variational problem, corresponding to the OC problem written by eq 4, we need to find explicit expressions for the control variables $u_j$. This can be achieved when it is possible to solve equations eq 4 for $u_i$ and then substitute it into eq 5. Then the problem of eq 5 subject to eq 4 could be rewritten as a pure variational problem in a similar way as for the mechanical case:

$$S=\int_{t_0}^{\tau}\Lambda(\xi_i,\dot{\xi}_i)dt\rightarrow\min\,,\ \xi_i(t_0)=\xi_{i0} \quad (7)$$

Now, for an isolated thermodynamic system at constant temperature and pressure ($T=const$ and $P=const$), when the thermodynamic potential $\Psi$ is Gibbs free energy $G(\xi_i)$ and the dissipation function after substitution of $u_i$ also became dependent on the extent coordinates $\xi_i$ ; $\Phi=\Phi(\xi_i,\dot{\xi}_i)$ in a general case, the thermodynamic Lagrangian will be

$$\Lambda(\xi_i,\dot{\xi}_i)=\Phi(\dot{\xi}_i,\xi_i)+G(\xi_i) \quad (8)$$

and the action functional to minimize:

$$S=\int_{t_0}^{\tau}(\Phi(\dot{\xi}_i,\xi_i)+G(\xi_i))dt\rightarrow\min\,,\ \xi_i(t_0)=\xi_{i0} \quad (9)$$

then the Euler-Lagrange equations become

$$\sum_{j=1}^{N}\frac{\partial^2\Phi}{\partial\dot{\xi}_j\partial\dot{\xi}_i}\ddot{\xi}_j+\sum_{j=1}^{N}\frac{\partial^2\Phi}{\partial\xi_j\partial\dot{\xi}_i}\dot{\xi}_j=\frac{\partial\Phi}{\partial\xi_i}+\frac{\partial G}{\partial\xi_i}\,,\ i,j=1,...,N\,. \quad (10)$$

In terms of generalized forces $X$ and fluxes $J$, using designations adopted in work:[24]

$$\frac{\partial G}{\partial\xi_i}=-X_i^G\,,\ \frac{d\xi_i}{dt}=J_i\,,\ \frac{d^2\xi_i}{dt^2}=\dot{J}_i\,;\text{ and }\frac{\partial\Phi}{\partial\xi_i}=-X_i^\Phi \quad (11)$$

the eq 10 could be rewritten as

$$\sum_{j=1}^{N}\alpha_{ij}\dot{J}_j+\sum_{j=1}^{N}\beta_{ij}J_j+X_i^\Phi+X_i^G=0\,,\ i,j=1,...,N \quad (12)$$

where $\alpha_{ij}=\dfrac{\partial^2\Phi}{\partial\dot\xi_j\partial\dot\xi_i}$ and $\beta_{ij}=\dfrac{\partial^2\Phi}{\partial\xi_j\partial\dot\xi_i}$ some coefficients from eq 10, $X_i^G$ – generalized chemical forces linked to thermodynamic potential $G$, $X_i^\Phi$ – generalized chemical forces, which are due to the explicit dependence of dissipative function $\Phi$ on the state variables.

However, because the open-end variational Lagrange problem is formulated, we also need to bear in mind the transversality condition (in our dissipative case – so-called natural boundary conditions, when at the end of relaxation $\xi_i(\tau)=0$ ):

$$\left(\Phi(\dot\xi_i,\xi_i)+G(\xi_i)-\sum_{i=1}^{N}\dot\xi_i\frac{\partial\Phi}{\partial\dot\xi_i}\right)_{\xi_i^*,\tau}=H(\xi_i^*,\dot\xi_i^*)=0 \cdot \tag{13}$$

This is obviously similar to the additional Pontryagin maximum principle demand for the optimal trajectories (suppose that $\xi^*$ gives a local minimum) of the equality of the Hamiltonian to zero $H(\xi^*,p^*)=0$ in an open-end OC problem. Eq 13 is, in fact, the energy conservation law for dissipative (thermodynamic) systems, which states that for the optimal trajectory the free energy $\Psi$ dissipated is equal to the energy dissipated by the mechanisms, formally contained in the dissipative function $\Phi$.

***Example 1.*** Let us follow the classical example mentioned above, when in the vicinity of the global equilibrium the dissipation function $\Phi$ can be written in a quadratic form as $2\Phi=R:\dot\xi\dot\xi$ where $\dot\xi$ is the vector of the generalized displacement derivatives, $R$ is the positive definite matrix. The quadratic approximation of thermodynamic potential can be written in the same way $2G=L:\xi\xi$ , where $\xi$ is the vector of generalized displacements, $L$ is the positive definite matrix. Then the dissipative Lagrangian can be written as a sum of two positive definite quadratic forms

$$\Lambda=\Phi+G=\frac{1}{2}R:\dot\xi\dot\xi+\frac{1}{2}L:\xi\xi \cdot \tag{14}$$

with the standard summation convention. Then the Euler-Lagrange equations are

$$\ddot\xi=R^{-1}L\xi \tag{15}$$

which describe the exponential relaxation due to the positive definite matrices $R$ and $L$. The general solution of these equations satisfying independent initial conditions can be written as:

$$\xi(t)=\xi_0\cosh(-\sqrt{R^{-1}L}t)+\dot\xi_0(\sqrt{R^{-1}L})^{-1}\sinh(-\sqrt{R^{-1}L}t), \tag{16}$$

where $\xi(t)$ displacement vector, $\xi_0=\xi_{t=t1}$, $\dot\xi_0=\dot\xi_{t=t1}$ are initial vectors. The decaying branch is $\xi(t)=\xi_0\exp(-\sqrt{R^{-1}L}t)$ and this branch requires $\dot\xi_0=-\sqrt{R^{-1}L})\xi_0$. Then it can be written the general flux-force relations as (if $R$ and $L$ are commutative)

$$J=-\sqrt{R^{-1}L}L^{-1}X, \tag{17}$$

in terms of generalized thermodynamic fluxes $J$ and generalized thermodynamic forces $X$ , which indicates the linear relations between the fluxes and forces in the vicinity of global equilibrium.

**Hamilton formulation.** Applying the Legendre transform to the thermodynamic Lagrangian from eq 8 we can obtain the thermodynamic Hamiltonian

$$H(\xi_i,p_i)=\sum_{i=1}^{N}\dot\xi_i(p_i)p_i-\Phi(\dot\xi_i(p_i),\xi_i)-G(\xi_i) \tag{18}$$

where $p_i$ are thermodynamic momenta. Then the canonical system can be written in terms of generalized forces $X$ and generalized fluxes $J$ as

$$J_i=\frac{\partial H}{\partial p_i}, \qquad \dot p_i=-\frac{\partial H}{\partial\xi_i}=X_i^\Phi+X_i^\Psi \tag{19}$$

Using the transversality condition we find for the optimal trajectory $\xi^*$, that $H(\xi_i^*,p_i^*)=0$

In *Example 1*, we can find the Hamiltonian using the Legendre transform when co-state variable $p$ is $p=R\dot\xi$ , then $\dot\xi=R^{-1}p$ and $\dot\xi^T=p^TR^{-1}$. Then the Hamiltonian is

$$H(\xi,p)=\frac{1}{2}R^{-1}:pp-\frac{1}{2}L:\xi\xi \tag{20}$$

and the canonical system looks:

$$\dot\xi=R^{-1}p, \qquad \dot p=L\xi \tag{21}$$

which could easily be transformed to eq 21, or in terms of flows and forces to $J=R^{-1}p,\dot p=-X$ ·

**Hamilton-Jacobi equations.** According to the Hamiltonian from eq 18 it is possible to obtain the Hamilton-Jacobi equation using the MED principle. In a general form the Hamilton-Jacobi equation could be written as

$$\frac{\partial S}{\partial t}+H(\xi_i,\frac{\partial S}{\partial\xi_i},t)=0 \tag{22}$$

where $S$ is the thermodynamic action and $H$ is the thermodynamic Hamiltonian that explicitly depends on time. When the thermodynamic Hamiltonian is time-independent this equation can be written as

$$H(\xi_i,\frac{\partial S_0}{\partial\xi_i})=E \tag{23}$$

where $S_0$ is the thermodynamic analogy of mechanical abbreviated action[39] $S_0=\int\sum_i p_i d\xi_i$ in an energetic representation, $E$ is a constant.

For *Example 1*, we can write the Hamilton-Jacobi equation in the form

$$\frac{1}{2}R^{-1}:\left(\frac{\partial S_0}{\partial\xi}\right)\left(\frac{\partial S_0}{\partial\xi}\right)-\frac{1}{2}L:\xi\xi=E \cdot \tag{24}$$

For the optimal trajectory $\xi^*$, $E=0$.

**General cooperative kinetics case.** We can now apply the formulated framework to study more general case in these terms, related to some well-studied nonlinear models/processes that are very characteristic of nonlinear kinetics and we can go on to compare the parameters obtained with classical linear process that gives exponential relaxation. For example, the logistic kinetics describes many cooperative processes in chemical reactions, in the denaturation of polymers, including proteins and DNA, ligand-receptor binding, metabolic networks and so on.

In contrast to the mechanical case of eq 1 let us suppose that the relationship between velocities/rates is not as simple as $\dot\xi=u$ , but is given by a system of equations that are linear relative to vector $u$:

$$\dot\xi=Fu \tag{25}$$

where $\xi$ and $u$ are the $N$-dimensional vectors, $\dot\xi$ is the time-derivative of vector $\xi$ , $F$ is a $N*N$ invertible matrix where coefficients are $f_{ij}=f_{ij}(\xi_1,\dots,\xi_N)$.

Let us also take the integrand (penalty) function $\Lambda$ for the OC problem (eq 5) in a quadratic form

$$\Lambda(\xi,u)=\frac{1}{2}K:uu+\frac{1}{2}L:\xi\xi \qquad (26)$$

where $K$ and $L$ are $N*N$ positive definite matrices.

We can reformulate the dynamic OC problem as a pure variational problem. For this we need to find $u$ from eq 25 and substitute it into eq 26. This means that matrix $F$ must be a nonsingular ($DetF \neq 0$) around a particular equilibrium state ($\xi=0$), and also throughout the part of state space traversed by the trajectory. Then $u=F^{-1}\dot{\xi}$ and $u^T=(F^{-1}\dot{\xi})^T=\dot{\xi}^T(F^{-1})^T$. Now we can rewrite the Lagrangian (eq 26) in terms of $\xi$ and $\dot{\xi}$, so

$$\Lambda(\xi,\dot{\xi})=\frac{1}{2}\dot{\xi}^T R\dot{\xi}+\frac{1}{2}\xi^T L\xi \qquad (27)$$

where we have designated $R(\xi)=(F^{-1})^T KF^{-1}$. Since derivatives $\frac{\partial\Lambda}{\partial\dot{\xi}_i}=R_{ij}\dot{\xi}_j$ and $\frac{\partial\Lambda}{\partial\xi_i}=\frac{1}{2}\frac{\partial R_{jk}}{\partial\xi_i}\dot{\xi}_j\dot{\xi}_k+(L\xi)_i$, and consequently $\frac{d}{dt}\frac{\partial\Lambda}{\partial\dot{\xi}_i}=R_{ij}\ddot{\xi}_j+\frac{\partial R_{jk}}{\partial\xi_k}\dot{\xi}_k\dot{\xi}_j$, finally, if matrix K is symmetric, then $R(\xi)=(F^{-1})^T KF^{-1}$ is also symmetric' we can obtain the Euler-Lagrange equations

$$R_{ij}\ddot{\xi}_j+\left(\frac{\partial R_{ij}}{\partial\xi_k}-\frac{1}{2}\frac{\partial R_{jk}}{\partial\xi_i}\right)\dot{\xi}_j\dot{\xi}_k=(L\xi)_i \quad , \qquad (28)$$

or equivalently,

$$R_{ij}\ddot{\xi}_j+\Gamma_{ijk}\dot{\xi}_j\dot{\xi}_k=(L\xi)_i\,, \qquad (29)$$

where $\Gamma_{ijk}=\frac{1}{2}\left(\frac{\partial R_{ij}}{\partial\xi_k}+\frac{\partial R_{ik}}{\partial\xi_j}-\frac{\partial R_{jk}}{\partial\xi_i}\right)$.

The correspondent to the Lagrangian from eq 27 Hamiltonian is

$$H(\xi,p)=2p^T(R^T+R)^{-1}R^T(R^T+R)^{-1T}p-\frac{1}{2}\xi^T L\xi=$$
$$=\frac{1}{2}p^T R^{-1}p-\frac{1}{2}\xi^T L\xi \qquad (30)$$

and canonical system will be

$$\begin{aligned}\dot{\xi}_i&=2(Z^T+Z)p_i\\ \dot{p}_i&=-2p^T(\partial Z/\partial\xi_i)p+(L\xi)_i\end{aligned} \quad , \qquad (31)$$

where $Z\equiv(R^T+R)^{-1}R^T(R^T+R)^{-1T}$. In the case where $R$ is symmetric and not dependent on the vector of displacements $\xi$, $\partial R/\partial\xi=0$ and eq 31 transforms to equation eq 21 for the linear case in *Example 1*.

Correspondent to eq 30 the Hamilton-Jacobi equation can be also written as

$$2\frac{\partial S_0}{\partial\xi}(R^T+R)^{-1}R^T(R^T+R)^{-1T}\frac{\partial S_0}{\partial\xi}-\frac{1}{2}\xi^T L\xi=E\,.$$

Or setting $p=\nabla S_0$ and taking into account that R is symmetric it can be written

$$\frac{1}{2}(\nabla S_0)^T R^{-1}\nabla S_0-\frac{1}{2}\xi^T L\xi=E\,. \qquad (32)$$

For the optimal (real) trajectory $\xi^*$, $E=0$.

**Example 2. Logistical cooperative model.** Logistical kinetics, characterised by symmetrical s-shaped relaxational curve, is typical for many cooperative processes. Let us consider one-dimensional scalar example and suggest that the eq 25 is $\dot{\xi}=f(\xi)u$, where $f(\xi)=1-\xi$, $(\xi<1)$. Then the Lagrangian for the problem according to eq 27 is

$$\Lambda(\xi,\dot{\xi})=\frac{r}{2}(1-\xi)^{-2}\dot{\xi}^2+\frac{l}{2}\xi^2 \qquad (33)$$

The scalar Euler-Lagrange equation becomes

$$r(1-\xi)\ddot{\xi}+r\dot{\xi}^2=l(1-\xi)^3\xi\,. \qquad (34)$$

Then by means of eq 33 the Hamiltonian can be found using the Legandre transform. We can also write the equation for optimal trajectory $\xi^*(t)$

$$\dot{\xi}^*=-\sqrt{\frac{l}{r}}(1-\xi^*)\xi^* \qquad (35)$$

which coincides in form with the well known logistical equation. Its solution is

$$\xi^*(t)=\left(\exp(-\sqrt{r^{-1}l}t)\right)\left(C+\exp(-\sqrt{r^{-1}l}t)\right)^{-1} \qquad (36)$$

and finally in thermodynamic terms the relation between the flux and force is

$$J=-\sqrt{r^{-1}l^{-1}}X+\sqrt{r^{-1}l^{-3}}X^2\,. \qquad (37)$$

One can see from this equation that in the vicinity of equilibrium ($\xi<<1$), when force are small, then second term becomes negligible comparably to first term, and this nonlinear expression for flux $J$ and force $X$ coincides with one-dimensional linear expression for exponential relaxation eq 17.

In addition, applying the Hamilton-Jacobi formulation eq 22-23, 32 to the one–dimensional logistic model from *Example 2* we can write the Hamilton-Jacobi equation for this scalar case as

$$\frac{(1-\xi)^2}{2r}\left(\frac{dS_0}{d\xi}\right)^2-l\frac{\xi^2}{2}=E \qquad (38)$$

where $S_0$ is so-called abbreviated action,[39] then we can obtain for the action $S$

$$S_0=\pm\sqrt{r}\int\left(\frac{\sqrt{2E+l\xi^2}}{1-\xi}\right)d\xi-Et \qquad (39)$$

the positive branch of eq 39 when $E=0$ leads to eq 35 with positive sign and is growing logistic branch. The second constant $\beta$ is

$$\beta=\frac{\partial S}{\partial\alpha}=\sqrt{r}\int\frac{d\xi}{(1-\xi)\sqrt{2E+l\xi^2}}-t \quad .$$

Finally we can obtain for optimal trajectory $\xi^*$ when $E=0$:

$$(t+\beta)\sqrt{\frac{l}{r}}=-\ln\frac{\xi^*}{1-\xi^*}\,, \qquad (40)$$

which, in fact, describes in slightly different way one-dimensional logistical kinetics, similarly to eq 36.

As mentioned in the introduction, cooperative phenomena are very diverse; they cover many fields from lasers and magnetism to social or metabolic networks. In this study an approach, which allows the use of a dynamic optimal control constraint system to build the variational framework for nonlinear chemical kinetics has been proposed, which is consistent with wider physics. The Lagrange/Hamilton and Hamilton-Jacobi formulations and general equations can be obtained and solved for particular cases. From the results, a general multi-dimensional linear case (which matches the classical linear relations between generalized thermodynamic fluxes and forces) demonstrates good applicability of this approach. A general nonlinear cooperative kinetics case is an example of the approximation/extension of the general non-linear model. This consideration narrows the gap between chemical thermodynamics, physics and other fields where cooperative phenomena are observed, it also creates a basis for the description of further classes of nonlinear processes.

## Conclusions

A simple framework has been introduced where the Lagrangian in a pure variational approach is built using the dynamic optimal control formulation. It is suggested that the transition to the pure variational approach can be carried out using analysis of possible regulatory mechanisms in the system, treating as the constraint system for the dynamic optimal control problem. In some cases, when the optimal control constrained system can be solved with the control variables, the corresponding variational problem can be formulated in a very explicit way. Then the Euler-Lagrange equations can be obtained, the corresponding Hamilton formulation can be found and related dissipative Hamilton-Jacobi equations can then be written in a general form. The cooperative kinetics case, important for many applications has been considered as a nonlinear example. In a case, when the constrained system in the dynamic optimal control problem is linear with regard to the control variables, the variational problem can be straightforward formulated. When the matrix of the dynamic constraint system has constant coefficients, and the thermodynamic potential has quadratic approximation (in the vicinity of equilibrium), this approach coincides with linear non-equilibrium thermodynamics. The expression in terms of generalized force and generalized flux has also been written for the logistical cooperative process, which appears as a quadratic approximation of the general nonlinear relation.

## References and Notes

(1) Englander, S. W.; Mayne, L.; Rumbley, J. N. *Biophys. Chem.* **2002**, *101-102*, 57.

(2) Qian, H. Biophys. J. **2008**, *95*, 10.

(3) Veytsman, B. A. *J. Phys. Chem.* **1993**, *97*, 7144.

(4) Sen, T. Z.; Feng, Y.; Garcia, J.V.; Kloczkowski, A.; Jernigan, R.L. *J. Chem. Theory Comput.* **2006**, *2*, 696.

(5) Licata, N.A.; Tkachenko, A.V. *Phys. Rev. Lett.* **2008**, *100*, 158102.

(6) Novak, I.L.; Slepchenko, B. M.; Mogilner, A.; Loew, L.M. *Phys. Rev. Lett.* **2004**, *93*, 268109.

(7) Sokolov, A.; Aranson, I. S.; Kessler, J. O.; and Goldstein, R. E. *Phys. Rev. Lett.* **2007**, *98*, 158102.

(8) Sum A.K.; Sandler, S.I. *J. Phys. Chem. A.* **2000**, *104*, 1121.

(9) Xantheas, S. S. *Chem. Phys.* **2000**, *258*, 225.

(10) Reich, L.; Weikl, T. R. *Proteins: Struct. Func. Bioinf.* **2006**, *63*, 1052.

(11) Gu, J.; Wang, J.; Leszczynski, J. *J. Phys. Chem. B* **2004**, *108*, 8017.

(12) Ludwig, R. *J. Mol. Liq.* **2000**, *84*, 65.

(13) Luty, T.; Yonemitsu, K. J. Phys. Soc. Jpn. 2004, 73, 1237.

(14) Bulsara, A.R.; Lindner, J.F.; In, V.; Kho, A.; Baglio, S.; Sacco, V.; Ando, B.; Longhini, P.; Palacios, A.; Rappel, W.-J. Phys. Lett. A. 2006, 353, 4.

(15) Haken, H. *Advanced Synergetics: Instability Hierarchies of Self-Organizing Systems and Devices;* Springer-Verlag: New York, 1993.

(16) Demishev, S.V.; Semeno, A. V.; Pronin, A. A.; Sluchanko, N. E.; Samarin, N. A.; Ohta, H.; Okubo, S.; Kimata, M.; Koyama, K.; Motokawa M.; and Kuznetsov, A. V. ***J. Supercond. Nov. Magnet.*** *2007***, 20,** ***105.***

(17) Hur, K. Le, Annals Phys. **2008**, *323*, 2208.

(18) Agarwal, G.S., *Phys. Rev. A.* **1978**, *18*, 1490.

(19) Agarwal, G.S.; Rattay, F., *Phys. Rev. A.* **1988**, *37*, 3351.

(20) Drobizhev, M.; Karotki, A.; Dzenis, Yu.; Rebane, A. *J. Phys. Chem. B* **2003**, *107*, 7540.

(21) Scott, J. *Social Network Analysis: A Handbook, 2nd ed.;* Sage: London, 2000.

(22) Varela, F.; Lachauz, J.P.; Rodriguez, E.; Martinerie, J. *Nat. Rev. Neurosci.* **2001**, *2*, 229.

(23) Demirel, Ya. *Int. J. Heat Mass Transf.* **2009**, *52*, 2018.

(24) Ziegler, H. *ZAMP* **1983**, *34*, 832.

(25) Paltridge, G.W. *Nature* **1979**, *279*, 630.

(26) Lorenz, R.D. *J. Non-Equilib. Thermodyn.* **2002**, *27*, 229.

(27) Kleidon, A.; Fraedrich, K.; Kunz, T.; Lunkeit, F. *Geophys. Res. Lett.* **2003**, *30*, 2223.

(28) Svoboda, J.; Turek, I.; Fischer, F.D. *Phil. Mag.* **2005**, *85*, 3699.

(29) Fratzl, P.; Fischer, F.D.; Svoboda, J. *Phys. Rev. Lett.* **2005**, *95*, 195702.

(30) Martyushev, L.M.; Seleznev, V.D. *Phys. Rep.* **2006**, *426*, 1.

(31) Moroz, A. *The Common Extremalities in Biology and Physics*; PHME: Minsk, 1997.

(32) Moroz, A. *Chem.Phys.Lett.* **2008**, *457*, 448.

(33) Dewar, R.C. *J. Phys. A: Math. Gen.* **2003**, *36*, 631.

(34) Dewar, R.C. *J. Phys. A: Math. Gen.* **2005**, *38*, L371.

(35) Prigogine, I. *Introduction to Non-Equilibrium Thermodynamics*; Wiley-Interscience: New York, 1962.

(36) Glansdorff, P.; Prigogine, I. *Thermodynamic Theory of Structure, Stability and fluctuations*; Wiley: New York, 1971.

(37) Gelfand, I.M.; Fomin, S.V. *Calculus of Variation*; Prentice-Hall, Englewood Cliffs: New Jersey, 1963.

(38) Pontryagin, L.S.; Boltyanskii, V.G.; Gamkrelidze, R.V.; Mischenko, E.F. *The mathematical Theory of Optimal Processes*; Interscience: New York, 1962.

(39) Landau, L.D.; Lifshitz, E.M. *Mechanics (3rd ed)*; Pergamon: London, 1976.